\documentclass[a4paper]{article}

\usepackage{icrc2013}
\usepackage[english]{babel}
\usepackage{url}

\title{The Aftermath of an Exceptional TeV Flare in the AGN Jet of IC 310}

\shorttitle{An Exceptional TeV Flare in the AGN Jet of IC 310}

\authors{
Dorit Eisenacher$^{1}$, Pierre Colin$^{2}$, Saverio Lombardi$^{3}$, Julian Sitarek$^{4}$, Fabio Zandanel$^{5,12}$, Francisco Prada$^{5}$, Elina Linfors$^{6}$, David Paneque$^{2}$, Dominik Els\"asser$^{1}$, Karl Mannheim$^{1}$, for the MAGIC Collaboration, 
Cornelia M\"uller$^{1,7}$, for the \textit{Fermi}/LAT Collaboration, Thomas Dauser$^{7}$, Felicia Krau\ss$^{7}$, Sven Wilbert$^{1}$, Matthias Kadler$^{1}$, J\"orn Wilms$^{7}$, Uwe Bach$^{8}$, Eduardo Ros$^{9,10,8}$, Talvikki Hovatta$^{11}$, for the OVRO team, 
Tuomas Savolainen$^{8}$, for the MOJAVE team}

\afiliations{
$^1$Universit\"at W\"urzburg, D-97074 W\"urzburg, Germany\\ 
$^2$Max-Planck-Institut f\"ur Physik, D-M\"unchen, Germany \\ 
$^3$INAF National Institute for Astrophysics, I-00136 Rome, Italy\\ 
$^4$IFAE, Edifici Cn., Campus UAB, E-08193 Bellaterra, Spain\\ 
$^5$Inst. de Astrof\'{\i}sica de Andaluc\'{\i}a (CSIC), E-18080 Granada, Spain\\
$^6$Tuorla Observatory, University of Turku, FI-21500 Piikki\"o, Finland\\ 
$^7$Dr. Remeis-Sternwarte Bamberg, Astronomisches Institut der Universit\"at Erlangen-N\"urnberg, ECAP, D-96049 Bamberg, Germany\\ 
$^8$Max-Planck-Institut f\"ur Radioastronomie, D-53121 Bonn, Germany\\
$^{9}$Observatori Astron\`omic, Universitat de Val\`encia, E-46980 Paterna,
Val\`encia, Spain\\
$^{10}$Departament d'Astronomia i Astrof\'{\i}sica, Universitat de
Val\`encia, E-46100 Burjassot, Val\`encia, Spain\\
$^{11}$Cahill Center for Astronomy \& Astrophysics, California Institute of Technology, Pasadena, CA 91125, USA\\
\scriptsize{$^{12}$ now at: GRAPPA Institute, University of Amsterdam, N-1098XH Amsterdam, Netherlands\\}
}

\email{Dorit.Eisenacher@astro.uni-wuerzburg.de}

\abstract{The nearby active galaxy IC\,310 ($z=0.019$), located in the Perseus cluster of galaxies is a bright and variable multi-wavelength emitter from the
radio regime up to very high gamma-ray energies above 100\,GeV. Very recently, a blazar-like compact radio jet has been found by parsec-scale VLBI imaging. 
Along with the unusually flat gamma-ray spectrum and variable high-energy emission, this suggests that IC 310 is the closest known blazar and therefore a key object for AGN research.\\
As part of an intense observing program at TeV energies with the MAGIC telescopes, an exceptionally bright flare of IC\,310 was detected in November 2012 reaching a flux level of up to 
$>0.5$\,Crab units above 300\,GeV. We have organized a multi-wavelength follow-up program, including the VLBA, Effelsberg 100\,m, KVA, \textit{Swift}, \textit{INTEGRAL}, 
\textit{Fermi}/LAT, and the MAGIC telescopes. \\
We present preliminary results from the multi-wavelength follow-up program with the focus on the response of the jet to this exceptional gamma-ray flare.}


\begin{document}
\maketitle

\section{Introduction}

Active galactic nuclei (AGN) harbor the most extreme objects in the Universe. They exhibit supermassive black holes with up to $10^{10}$ solar masses surrounded by an
accretion disk and a dust torus. Perpendicular to the plane of the disk and the torus, collimated relativistic plasma outflows are ejected away from the AGN. Particular AGN are the so-called blazars for which the angle between those jets 
and the line-of-sight is believed to be small. Those objects are dominated by non-thermal emission observed at all wavelengths of the 
electromagnetic spectrum. They are characterized by rapid variations of their flux seen in all energy bands from radio to gamma-ray. 
In contrast, radio galaxies are AGN where the angle to the line-of-sight is large.

Most of the AGN detected up to the very high energy (VHE) regime ($>$ 100\,GeV) belong to the class of blazars. 
So far only three objects, M\,87, Centaurus\,A, and NGC\,1275 which are radio galaxies were discovered to be VHE emitters as well. 

%In particular, extremely fast variability of the flux is known for TeV emission as reported for rapid flares observed form Mrk\,501 
%and PKS\,2155-304 with durations of only a few minutes (Albert 2007; Aharonian 2007). These observational results are 
%challenging for the current blazar models by limiting the size of emission region to be smaller than the gravitational radius of
%the black hole.

The nearby ($z = 0.019$) lenticular (S0) galaxy IC\,310 exhibits an AGN, which has been detected at gamma-ray energies above 30\,GeV with 
\textit{Fermi}/LAT \cite{neronov10} as well as above 260\,GeV with the MAGIC telescopes \cite{aleksic10, aleksic13}.
The measured spectrum in the gamma-ray band is very hard.

In the past, the object was thought to be a head-tail radio galaxy \cite{sijbring}. 
Those are typically found in clusters of galaxies and are characterized by their extended jets which are pointing in the direction determined by the galaxy's motion through the intra-cluster medium (ICM).
In contrast to this classification, a milliarcsecond-scale one-sided blazar-like jet oriented along the same position angle as the 
kiloparsec scale radio structure has been found for IC\,310 at 8.4\,GHz \cite{kadler12} and confirmed by MOJAVE observations at 15\,GHz (see Fig.~\ref{VLBA}). Assuming that the one-sidedness is due to Doppler boosting, the angle between
the jet axis of the AGN and the line-of-sight has been estimated to be $<38^{\circ}$ \cite{kadler12}. The stability of the jet orientation from parsec to kiloparsec 
scales in IC\,310 argues against its classification as a head-tail radio galaxy, i.e., 
there is no indication of an interaction with the intracluster medium that would determine the direction of the tail. 

In \cite{rector99, kadler12} the authors suggested already that IC\,310 belongs to a transitional population between BL Lac and FR~I (i.e. low power) radio galaxy based on the optical,
radio and X-ray observational results of the source. The weak optical emission lines of IC\,310 are typically found for FR~I radio galaxies
but its radio to X-ray non-thermal continuum is similar to low-luminosity BL~Lac objects \cite{owen96}. Observations
by \textit{XMM-Newton} \cite{sato} showed the X-ray emission to be dominated by nuclear
emission with a featureless power-law X-ray spectrum that is typical for a BL~Lac-type object, as well. 
\textit{Chandra} observations show a faint X-ray halo extending in the direction of the radio tail \cite{dunn}.

The observational data, showing daily TeV flux variability and strong variability in the X-ray spectra and flux, suggest that 
IC\,310 is either the closest known blazar as already suggested by \cite{kadler12, aleksic13} or one of four VHE loud radio galaxies.

\section{Observation Campaign and Results}

\begin{figure}[!t]
 \vspace{5mm}
 \centering
 \includegraphics[width=3.in]{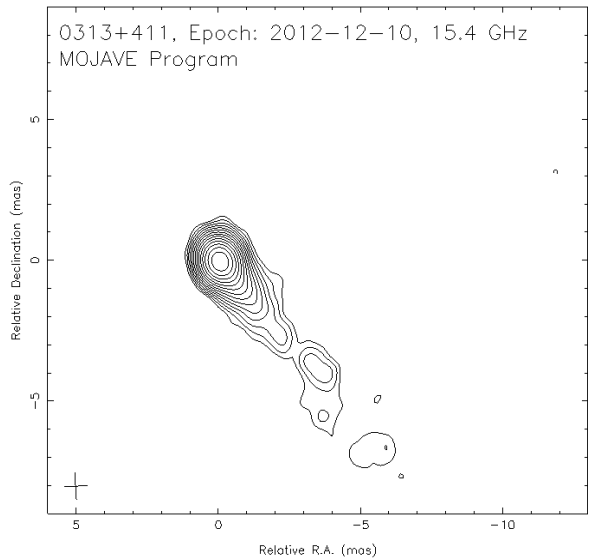}
 \caption{Parsec-scale jet of IC\,310 as measured with the VLBA at 15\,GHz on December 12, 2012 as part of the MOJAVE project. The lowest contour and the brightness peak are 0.17\,mJy\,beam$^{-1}$ and 73.0\,mJy\,beam$^{-1}$. 
The beam has a size of 0.98$\times$0.82\,mas at 3.5$^{\circ}$. The image has been taken from \url{http://www.physics.purdue.edu}. }
 \label{VLBA}
 \end{figure}

\begin{figure}[!t]
 \vspace{5mm}
 \centering
 \includegraphics[width=3.in]{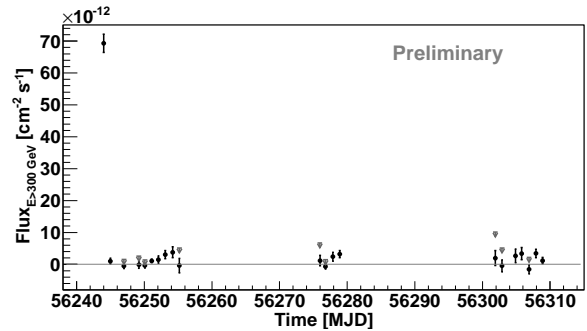}
 \caption{Light curve of IC\,310 measured with MAGIC above 300\,GeV between November 2012 to January 2013. The upper limits were calculated applying the method 
by \cite{rolke}, using a confidence level of 95\% and assuming 30\% systematic uncertainty.}
 \label{Lightcurve}
 \end{figure}

The first successful multi-wavelength campaign on IC\,310 was started in autumn 2012 after high resolution Very Long Baseline Interferometry
(VLBI) was conducted with the European VLBI Network.  

For this campaign, VHE observations of IC\,310 were carried out between November 2012 and January 2013 with the MAGIC telescopes which are two Imaging Atmospheric Cherenkov Telescopes
located on the island of La Palma at an altitude of 2200\,m.
Both telescopes consist of a mirror dish of 17\,m diameter associated with a fast imaging camera of 3.5$^{\circ}$ field-of-view.
In summer 2011 to autumn 2012 the MAGIC telescopes underwent a major upgrade in order to assure a stable performance and operation of the telescopes
for the upcoming years \cite{mazin}. The upgrade included a replacement of the readout system of both telescopes with a digitizing system based on the DRS-4 chip
and a new MAGIC-I camera, a clone of the MAGIC-II camera, equipped with 1039 channels of 0.1$^{\circ}$ pixels, increasing the trigger region.

%The data of the Perseus cluster from November 2012 to February 2013 were taken in two different so-called wobble modes since two VHE objects were 
%already detected by MAGIC in the field-of-view of the Perseus cluster. In the first mode the telescopes are 
%pointing alternately to four sky positions with an offset of 0.4$^{\circ}$ away from the center of the cluster, i.e. NGC\,1275. The individual wobble positions 
%were chosen in a way that two of the four wobble positions are 0.4$^{\circ}$ away form IC\,310 as well. The remaining ones have an offset of $0.938^{\circ}$ from the object.
%In the second case the wobble center is chosen to be in between NGC\,1275 and IC\,310 with the sky coordinates of R.A. 03h\,18m\,25s and Decl. 41$^\circ$\,25$^\prime$\,05$^{\prime\prime}$. 
%The wobble offsets are then chosen to be 0.26$^{\circ}$ in order to guarantee a standard 0.4$^{\circ}$ distance to the individual objects.

Optical observations were performed with the Kungliga Vetenskapsakademien (KVA) 35\,cm telescope on La Palma that operates in close collaboration
with the MAGIC telescopes as well as with optical telescopes at the Observatoire de Haute Provence (France). 
Additionally, further observations in the radio band were conducted by the Effelsberg 100\,m telescope and data from the OVRO blazar monitoring program at 15\,GHz have been used for this campaign \cite{richards}. 
VLBI monitoring observations at 15\,GHz were commenced as part of the
MOJAVE project \cite{lister}, see also Fig.~\ref{VLBA}. The UV and the X-ray coverage was provided by Target of Opportunity 
observations organized with \textit{Swift}. The hard X-ray and soft gamma-ray regime has been covered by \textit{Swift}/BAT, \textit{INTEGRAL} and \textit{Fermi}/LAT. 

\begin{figure*}[!t]
  %\vspace{5mm}
  \centering
  \includegraphics[width=6.in]{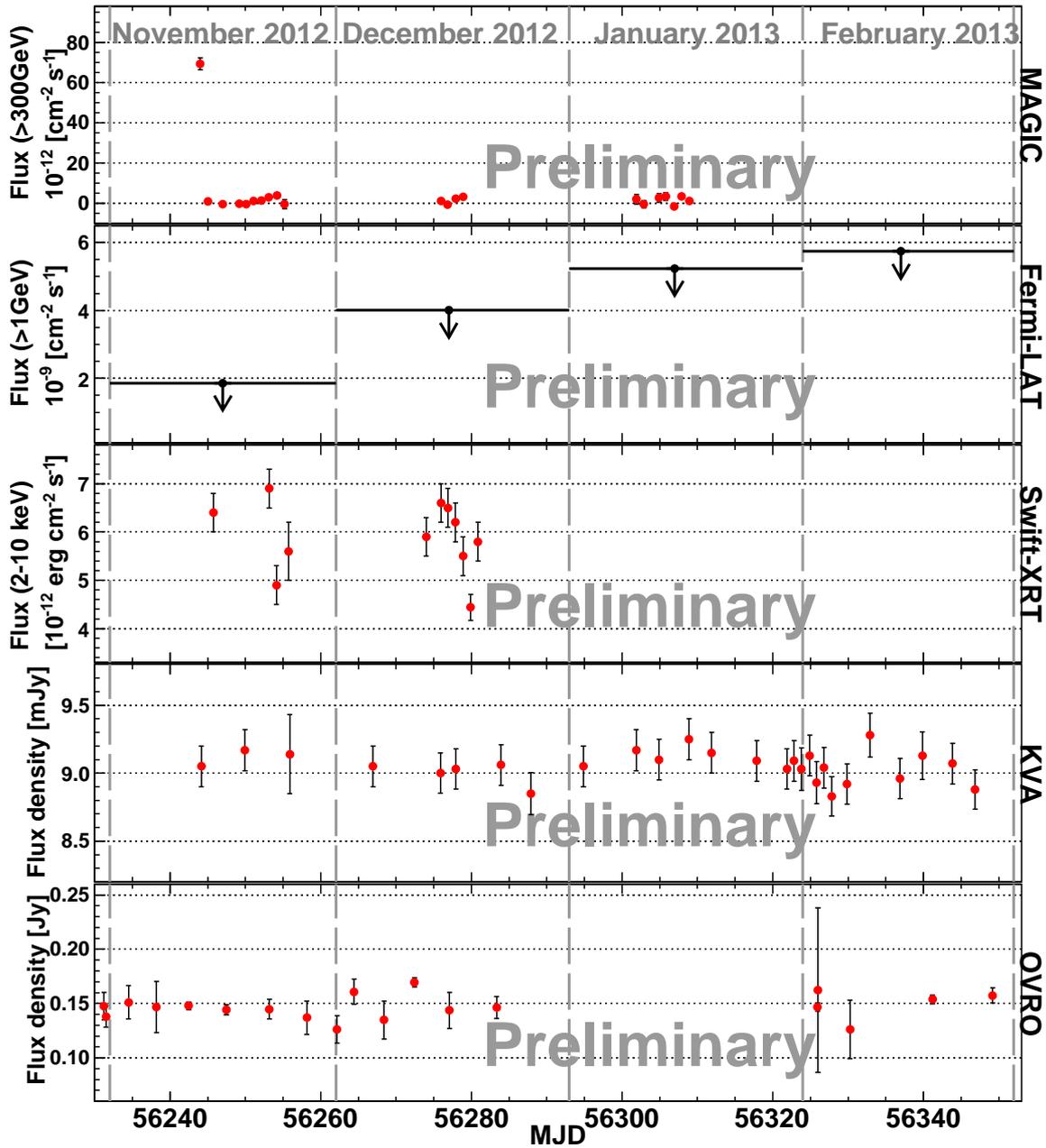}
  \caption{Multi-wavelength light curve of IC\,310. \textit{Top to bottom panel:} MAGIC above 300\,GeV, \textit{Fermi}-LAT above 1\,GeV, \textit{Swift}-XRT, KVA R-band data and, OVRO at 15\,GHz from November 2012 to February 2013. 
The optical data from KVA presented here are not host-galaxy corrected.}
  \label{LightcurveMWL}
 \end{figure*}

\begin{figure}[!t]
\vspace{5mm}
  \centering
  \includegraphics[width=3.in]{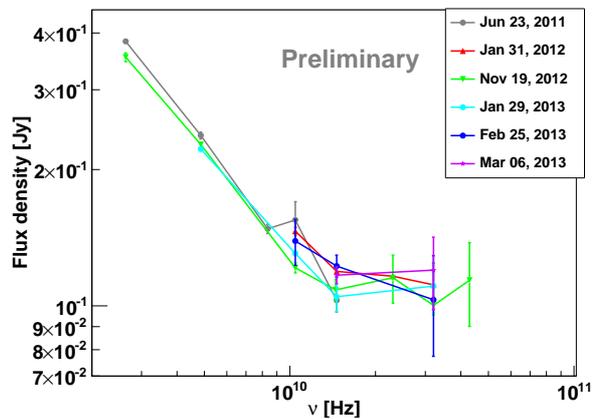}
  \caption{Flux-density measurements with the Effelsberg 100-m telescope as a function of frequency at different epochs. As of March 6, 2013, no enhancement of the flux density at
the highest frequencies after the TeV flare has been observed.}
  \label{Effelsberg}
 \end{figure}

The preliminary light curve measured by MAGIC above 300\,GeV is shown in Fig.~\ref{Lightcurve}, and comprised with multi-wavelength data in Fig.~\ref{LightcurveMWL}. The analysis of the \textit{INTEGRAL} data is in progress. 

During this campaign MAGIC observations carried out on MJD 56243.9-56244.1 detected the highest flare ever detected from IC\,310 in the VHE range \cite{MAGICATel}. 
The analysis resulted in a detection with the significance of $\sim$25 standard deviations ($\sigma$). 
The mean flux above 300\,GeV is 56\% of the Crab Nebula flux (C.U.). During previous observations in 2009/2010 this mean flux was measured to be $(2.5\pm0.4)$\% C.U. \cite{aleksic10}
and $(12.8\pm0.1)$\% C.U. for the high state flux reported in \cite{aleksic13}.

Quasi-simultaneous observations of IC\,310 after the VHE flare were performed on MJD 56245.67 with \textit{Swift} \cite{SwiftATel}. The spectrum of IC\,310 was observed to 
be similar to an earlier \textit{Swift}-observation in January 2012, but a factor of 1.5 brighter. Based on the \textit{Swift} quick-look-data, the count 
rate was 0.24\,counts/s (0.5--10\,keV). The spectrum can be well described by an absorbed power law with a photon index of $1.94\pm0.16$ and $N_{\mathrm{H}}=(1.3\,\pm\,0.4)\times10^{21}$\,cm$^{-2}$, 
corresponding to a 2--10\,keV flux of $(6.4\pm0.8)\times\,10^{-12}$\,cgs. The measured 
$N_{\mathrm{H}}$ 
value is consistent with the Galactic 21\,cm $N_{\mathrm{H}}$ as obtained from the Leiden-Argentine-Bonn survey \cite{kalberla}. 
In January 2012, the photon index was $2.14\pm0.12$. Compared to the February 26, 2003 \textit{XMM-Newton} observations of IC\,310 \cite{sato} the source flux was enhanced by a factor of 4.6 and the source spectrum was significantly harder (\textit{XMM}-Value: 2.5). 
Further \textit{Swift} observations during the campaign show a hint for variability of the flux of the order of $\sim3\sigma$ in the 2--10\,keV band.

After the VHE flare the UV flux observed by \textit{Swift}-UVOT is virtually unchanged compared to previous observations and appears to be dominated by the host 
galaxy. The same behavior is seen for the optical data from KVA.

The radio data from OVRO at 15\,GHz show no significant variability. The radio spectra obtained from Effelsberg show no large variations (see Fig.~\ref{Effelsberg}).

\section{Discussion and Conclusion}

We present preliminary results from the first multi-wavelength campaign of the TeV loud object IC\,310.
We interpret the variable emission from X-ray to VHE to originate from the central blazar-like engine rather than being produced by interactions with the ICM.
During those observations, an exceptional TeV flare has been detected with MAGIC.

Several authors have reported a coincidence between the ejection of new radio components in VLBI maps and high gamma-ray 
flux periods from different blazars \cite{jorstad, marscher, kovalev}. The typical timescale between a gamma-ray flare and the ejection are between less than one month and several months. 
The VHE flare observed from IC\,310 in November 2012 gives the opportunity to study in detail 
the response of an individual radio jet to an unprecedented VHE flare. As IC\,310 is included in the MOJAVE program the pc-scale jet has been monitored.
Additionally, the radio spectrum of IC\,310 has been observed with the Effelsberg telescope 
to find evidence for such an ejection. Especially at higher frequencies, e.g., above 10\,GHz where the radio emission is dominated by a flat blazar-like spectrum (see Fig.~\ref{Effelsberg}), 
any enhancement of the flux density could indicate a new radio component appearing in the jet. Below 10\,GHz the spectrum is steep, indicating extended, optically thin emission. 
As of February 2013, the radio monitoring did not show any significant response of the radio jet.

A more detailed analysis of the extensive amount of multi-wavelength data is currently ongoing. 

\vspace*{0.5cm}
\footnotesize{{\bf Acknowledgment: }{We would like to thank the Instituto de Astrof\'{\i}sica de
Canarias for the excellent working conditions at the
Observatorio del Roque de los Muchachos in La Palma.
The support of the German BMBF and MPG, the Italian INFN, 
the Swiss National Fund SNF, and the Spanish MICINN is 
gratefully acknowledged. This work was also supported by the CPAN CSD2007-00042 and MultiDark
CSD2009-00064 projects of the Spanish Consolider-Ingenio 2010
programme, by grant DO02-353 of the Bulgarian NSF, by grant 127740 of 
the Academy of Finland,
by the DFG Cluster of Excellence ``Origin and Structure of the 
Universe'', by the DFG Collaborative Research Centers SFB823/C4 and SFB876/C3,
and by the Polish MNiSzW grant 745/N-HESS-MAGIC/2010/0.\\
The Fermi-LAT Collaboration acknowledges support from a number of agencies and institutes for both development and the operation of the LAT as well as scientific data analysis. These include NASA and DOE in the United
States, CEA/Irfu and IN2P3/CNRS in France, ASI and INFN in Italy, MEXT,KEK, and JAXA in Japan, and the K. A. Wallenberg Foundation, the Swedish
Research Council and the National Space Board in Sweden. Additional support from INAF in Italy and CNES in France for science analysis during the operations phase is also gratefully acknowledgeged.\\
This research has made use of data obtained from the High Energy Astrophysics Science Archive Research Center (HEASARC), provided by NASA's Goddard Space Flight Center.\\
This work is based on observations with the 100-m telescope of the MPIfR (Max-Planck-Institut für Radioastronomie) at Effelsberg.\\
This research has made use of data from the MOJAVE database that is maintained by the MOJAVE team (Lister et al., 2009, AJ, 137, 3718).  
The Very Long Baseline Array (VLBA) is operated by the National Radio Astronomy Observatory, a facility of the National Science Foundation operated under cooperative agreement by Associated Universities, Inc.\\
The OVRO 40 M Telescope Fermi Blazar Monitoring Program is supported by NASA under awards NNX08AW31G and NNX11A043G, and by the NSF under awards AST-0808050 and AST-1109911.\\
This work made use of the Swinburne University of Technology software
correlator (Deller et al., 2011, PASP, 123, 275), developed as part of the Australian 
Major National Research Facilities Programme and operated under licence.\\
E. R. was partially supported by the Spanish MINECO projects AYA2009-13036-C02-02 and AYA2012-38491-C02-01 and by the Generalitat Valenciana project PROMETEO/2009/104, as well as by the COST MP0905 action `Black Holes in a Violent Universe'.}}

\end{document}